# Graphene under direct compression: stress effects and interlayer coupling

**Elena del Corro**[*,1], **Miriam Peña-Álvarez**[1], **Michal Mračko**[2], **Radek Kolman**[2], **Martin Kalbáč**[1], **Ladislav Kavan**[1], **Otakar Frank**[1]

J.Heyrovsky Institute of Physical Chemistry of the AS CR v.v.i., Dolejskova 2155/3, CZ 182 23 Prague 8, Czech Republic

Institute of Thermomechanics of the AS CR, v.v.i., Dolejskova 1402/5, CZ 182 23 Prague 8, Czech Republic

In this work we explore mechanical properties of graphene samples of variable thickness. For this purpose, we coupled a high pressure sapphire anvil cell to a micro-Raman spectrometer. From the evolution of the G band frequency with stress we document the importance the substrate has on the mechanical response of graphene. On the other hand, the appearance of disorder as a consequence of the stress treatment has a negligible effect on the high stress behaviour of graphene.

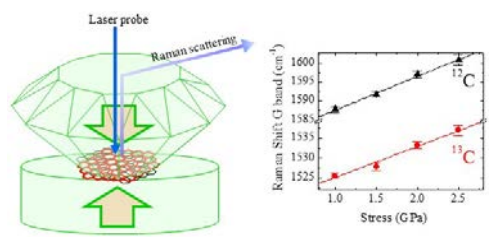

Isotopic labelled twisted bilayer graphene under high compression characterized by Raman spectroscopy.

**1 Introduction** Graphene has attracted great interest in the last decade due to its unique structure [1] which provides its fascinating mechanical and electrical properties [2]. Graphene is a zero gap semiconductor which also presents the greatest stiffness found in nature. Moreover, strain engineering of graphene has become a promising route for tailoring its electronic properties [3, 4]. Several approaches have been followed in order to induce strain in graphene [5-12], among which we can find high pressure experiments [9-12]. In most of these studies Raman spectroscopy is chosen as the main characterization technique since it represents a non-destructive tool for in-situ strain sensing, which also allows to address doping effects [13, 14]. Concerning the high pressure experiments, we find several studies of graphene in the literature, including samples prepared by exfoliation on Si/SiO$_2$ [9, 10] and by chemical vapour deposition (CVD) on copper [11]. All of these studies have been performed under hydrostatic conditions, using various pressure-transmitting media, to assure the same stress acting along all directions. Furthermore, some studies have been reported for suspended few-layers graphene flakes [12]. A comprehensive analysis of the mentioned previous studies reveals that the compressibility of the substrate plays a key role in the mechanical response of graphene. In contrast, doping phenomena have no influence on the mechanical response of graphene, and do not affect the strain coefficients showed by the G Raman peaks of graphene [11]. In this work we perform direct out-of-plane compression on exfoliated and CVD graphene samples, all supported on sapphire discs. We study graphene samples of different thickness, ranging from mono (1L) to trilayer (3L). Additionally, we isotopically label graphene, so that the behaviour of individual layers can be addressed [15], shedding some light on the graphene-substrate and graphene-graphene interaction under compression.

**2 Methods** A set of graphene samples with different number of layers (1 to 3) were prepared by mainly two methods: mechanical exfoliation and CVD. Specifically, the labelled twisted bilayer graphene (tBLG) were prepared by CVD as single layer graphene samples of $^{12}$C and

$^{13}$C (as described elsewhere [16]) and then sequentially transferred to a sapphire disc by the common PMMA transfer method [17]. Atomic Force Microscopy (AFM) images of the transferred layer is shown in Figure S1 (Supporting Information). Single and bilayer graphene exfoliated samples on Si/SiO$_2$, with lateral dimensions of about 20 μm, were transferred to a sapphire disc by a dry transfer method [18]. The experimental setup consists of a gem anvil cell coupled to a Raman spectrometer LabRAM HR from Horiba Jobin-Yvon. Samples were excited with an Ar/Kr laser working at 488.0 nm, keeping the power below 1 mW in order to avoid sample heating. We used a 50x objective which provides a laser spot on the sample of about 2 μm in diameter. A 600 grooves mm$^{-1}$ grating provided a spectral resolution of ~1.8 cm$^{-1}$. In order to perform direct out-of-plane compression we use a modified sapphire cell where one anvil (~350 μm diameter culet) is opposed to a disc (2 mm thick x 10 mm in diameter) containing the sample. See Supporting Information for the alignment and stress distribution in the high pressure cell, calculated by Finite Elements (FE) method. In such conditions, the use of conventional stress marker is inadequate and the stress is estimated from the evolution of the Raman features of sapphire [19]. At each stress step single spectra or Raman maps were registered in exfoliated or CVD samples, respectively.

**3 Results** Sapphire-supported exfoliated bilayer graphene (see Methods) was subjected to direct out-of-plane compression, up to 4.5 GPa, and characterized with Raman spectroscopy; Figure 1 shows the evolution of the Raman spectrum with increasing non-hydrostatic stress. We present the spectral region from 1200 to 3000 cm$^{-1}$ which includes the most intense Raman features of graphene, the D, G and 2D bands. As expected, in comparison with graphite compressed under identical conditions [20], the Raman spectrum blue-shifts and broadens with stress.

The same high stress experiments (up to 2.5 GPa) were performed on isotopically labelled tBLG samples. At each stress step of 0.5 GPa a Raman map (40x40 μm) in the same region of the sample was recorded. Within the Raman maps two areas can be distinguished: a 400 μm$^2$ crystal grain with enhanced G band (corresponding to a critical twist angle of ~13° [21]); and the rest of the sample corresponding to a random region (twist angle different from 13°), thus with the 2D band as the most intense Raman contribution. In Figure 2 we present selected Raman spectra measured in the same sample spot of the latter random region. In labelled tBLG we can distinguish two D, G, and 2D bands, originating from vibrations of the particular isotope [15]. The phonon frequency is inversely proportional to the atomic mass, therefore the Raman bands at lower frequency correspond to the $^{13}$C isotope layer. Such a differentiation is highly advantageous since we can distinguish the effect of increasing stress on each layer and evaluate coupling effects in the graphene layers, between each other and with the substrate. In analogy to exfoliated BLG, the spectrum of labelled CVD tBLG upshifts and broadens with stress.

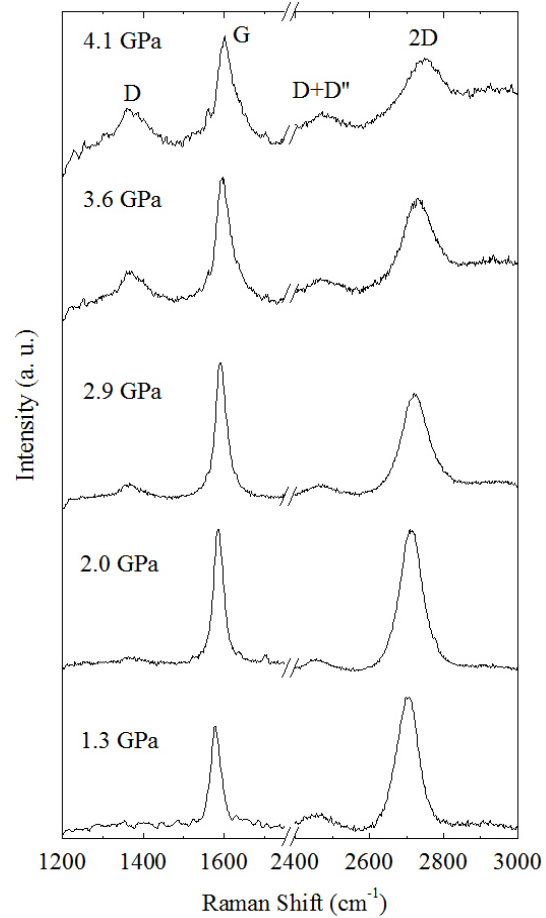

**Figure 1** Selected Raman spectra of exfoliated BLG with increasing compression.

For both types of bilayer samples, i.e., exfoliated and CVD-grown, an increase of the intensity of defect related bands (D and D´) is observed. Such increase of disorder upon compression was observed in graphite and was related to the appearance of shear stresses in the anvils [22]. A detailed analysis of the defects generation in the different samples, exfoliated and CVD, is shown later.

For the analysis of the stress response of the different graphene samples studied in this work we chose primarily the G band, since especially in bilayer samples, the 2D band could be prone to frequency changes depending on the twist angle [21]. Additionally, the Raman shift of the G band as a function of stress is usually employed as stress sensor and more data are available for the comparison [23].

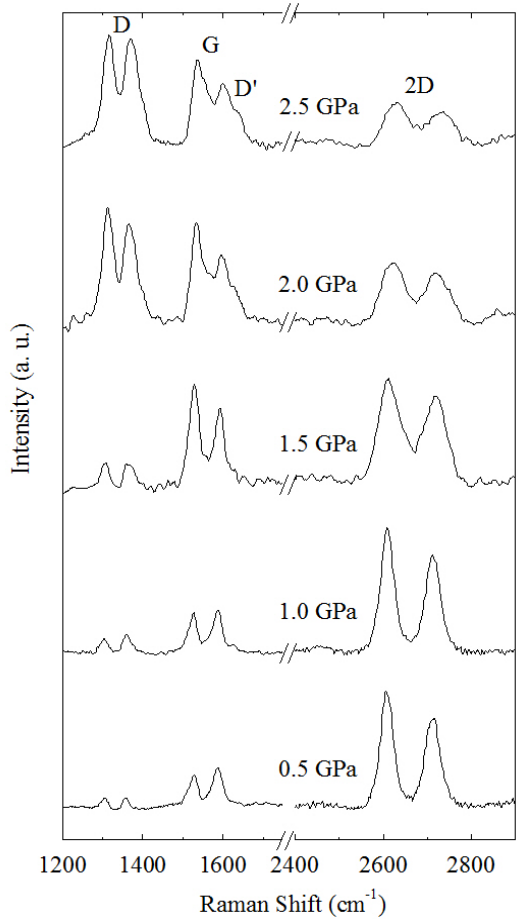

**Figure 2.** Selected Raman spectra of CVD labelled tBLG with increasing compression.

In analogy to exfoliated BLG, the spectrum of labelled CVD tBLG upshifts and broadens with stress. For the analysis of the stress response of the different graphene samples studied in this work we chose primarily the G band, since especially in bilayer samples, the 2D band could be prone to frequency changes depending on the twist angle [21]. Additionally, the Raman shift of the G band as a function of stress is usually employed as stress sensor and more data are available for the comparison [23].

We present the Raman shift of the G band with stress in Figure 3, for exfoliated BLG and for both graphene sheets in labelled tBLG, and the stress coefficients of the G band ($\delta\omega_G/\delta\sigma$) for each case are reported in Table 1. We obtain a similar coefficient, within the confidence interval (CI), for the exfoliated BLG and the both layers in labelled tBLG. Despite $^{12}$C slope seems slightly larger than that of $^{13}$C, we can assume the same stress coefficient for both graphene layers, given the instrumental resolution, since the frequency difference $\omega_{12C}-\omega_{13C}$ remains constant with stress at $63\pm1$ cm$^{-1}$.

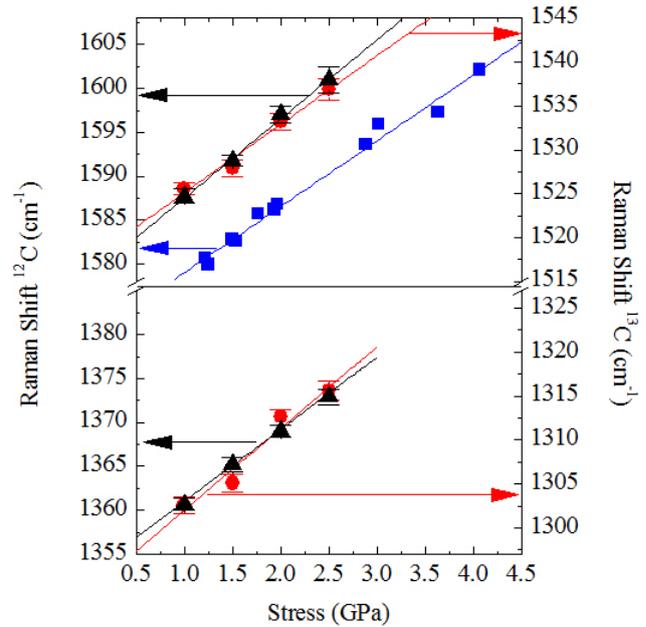

**Figure 3** Raman shift of the G and D bands as a function of stress for exfoliated BLG (blue squares) and for both graphene sheets in labelled tBLG (red circles and black triangles for the $^{13}$C top and $^{12}$C bottom layers, respectively). In the labelled tBLG sample, the error bars indicate the standard deviation of the Raman shift averages obtained from Raman maps registered at each stress step.

**Table 1** Stress coefficients ($\delta\omega_G/\delta\sigma$) and 95% confidence interval (CI) for the G and D bands, in cm$^{-1}$/GPa. Frequency correlation between G and 2D band ($\delta_{2D}/\delta_G$).

|  | $\delta\omega_G/\delta\sigma$ (CI) | $\delta\omega_{2D}/\delta\omega_G$(CI) | $\delta\omega_D/\delta\sigma$ (CI) |
|---|---|---|---|
| $^{12}$C (bottom) | 9.1 (4.0) | 2.0 (0.3) | 8.2 (2.9) |
| $^{13}$C (top) | 7.6 (3.5) | 2.1 (0.6) | 9.3 (3.8) |
| exfoliated | 7.5 (0.7) | 1.8 (0.2) | - |

The expected pressure coefficient of the G band, $\delta\omega_G/\delta\sigma$, of a suspended graphene layer under hydrostatic conditions is about 5-6 cm$^{-1}$/GPa [12]; however, we can find larger reported coefficients for supported samples on different substrates (Si/SiO$_2$ or Cu) [10,11]. Additionally, for the same supported kind of sample, different stress coefficients can be found depending on the pressure transmitting medium (PTM), from argon to alcohols, used to achieve the hydrostatic conditions, see Table 2. The observed difference in the G band stress coefficient was at first wrongly attributed to the coexistence of doping and stress effects in the experiments. It is well known that the peak position of the G and 2D bands of graphene is affected by the type and amount of doping in the sample [13]; however, the pressure coefficient of the Raman bands position is not affected by the initial or pressure-induced dop-ing effects. Moreover, doping and strain effects can be distinguished using the 2D to G frequency correlation ($\delta\omega_{2D}/\delta\omega_G$), so that when only mechanical effects are present such correlation is expected to be 2.2 [13]. For both our experiments, the $\delta\omega_{2D}/\delta\omega_G$ slope is slightly lower than 2.2, thus indicating some doping effects, probably due to the presence of some remnant polymer or glue from the transfer or exfoliation process, respectively (see Figure S1, Supporting Information). However, as commented above, this fact does not affect the value of $\delta\omega_G/\delta\sigma$, and, by comparison with data in Table 2, we conclude that the stress response of graphene under direct out-of-plane compression is comparable to that under hydrostatic pressure (in supported samples).

**Table 2** Reported pressure slopes of the G Raman band of graphene in hydrostatic compression, $\delta\omega_G/\delta P$, (in cm$^{-1}$/GPa) for suspended and different supported samples and various PTM.

| Reference | $\delta\omega/\delta P$ | layers | support | PTM |
|---|---|---|---|---|
| Proctor et al. [9] | 5.0 | few | --- | nitrogen |
| Nicolle et al [10] | 7.6 – 10.5 | 1 | Si/SiO$_2$ | Ar-alcohol |
| Filintoglou et al. [11] | 9.2 – 5.6 | 1 (CVD) | Cu | alcohol |
| Soldatov et al. [12] | 5.6 – 5.9 | 1 | --- | alcohol |

The different $\delta\omega_G/\delta P$ found in the literature for supported graphene samples are understood taking into account two factors. First, the stress response of the substrate, since the compressibility of the substrate may affect the stress transfer to the graphene layer, as well as the adherence between sample and substrate [11]. And second, the interaction between graphene and the pressure transmitting medium could affect $\delta\omega_G/\delta P$, as it may increase under pressure and become as large as the graphene-substrate adherence, leading to a pressure response similar to that of suspended graphene (5.6 cm$^{-1}$/GPa) [12]. In our experiments, we do not use pressure media and the samples (all of them are supported by a sapphire disc) are subjected to out-of-plane compression along the perpendicular direction. For this reason, we should expect always a stress coefficient larger than for suspended graphene, since the sample is always sandwiched between sapphires and cannot experience any detachment during the compression process. In agreement with that, our reported stress coefficients for the G band are in all cases larger than 5.6 cm$^{-1}$/GPa. The stress behaviour of the D band could be studied thanks to the use of sapphire anvil alternative to the common diamond ones. Raman shift of the D band as a function of stress is shown in Figure 3 and the corresponding stress coefficient are summarized in Table 1. We observe a similar stress coefficient for the top and bottom layer, in both cases about one half of the 2D band coefficient.

The stress coefficients reported in Table 1 were obtained in the stress range starting at 1 GPa. The reason for the higher onset of the fitting region is that the behaviour of $\delta\omega_G/\delta\sigma$ below 1 GPa, i.e. in the first stress step, differs from a linear evolution. We found an anomalous shift of the G band at the first stage of compression, not reported before. In order to further analyse this phenomenon, we carried out a comparative study of graphene samples with 1 to 3 layers and compressed them in a low stress regime, up to 1 GPa. In Table 3 we present the Raman shift of the G band when the cell is closed (0.5 GPa) and loaded (1.0 GPa), for different types of samples.

**Table 3** Average Raman shift of the G band, $\omega_G$, (standard deviation) at the first stages of compression, 0.5 and 1.0 GPa, and frequency difference, $\Delta\omega_G$, for various samples with increasing number of graphene layers (all in cm$^{-1}$)

| sample | $\omega_{G/0.5\,GPa}$ | $\omega_{G/1.0\,GPa}$ | $\Delta\omega_G$ |
|---|---|---|---|
| 1L exfoliated | 1581.3 (0.5) | 1581.2 (0.5) | –0.1 |
| 1L CVD | 1581.3 (0.5) | 1581.1 (0.5) | –0.2 |
| 2L exfoliated | 1582.6 | 1581.5 | –1.1 |
| 2L CVD | 1593.0 (1.01) | 1587.7 (0.8) | –5.3 |
| 3L CVD | 1592.2 (2.0) | 1585.4 (1.0) | –6.8 |

Interestingly, in contrast to what one could expect, the G band of graphene downshifts under compression from 0.5 to 1 GPa, for samples with more than one layer, finding the larger downshift in the thicker sample (3L). Moreover, for the same number of layers, such downshift is larger for CVD samples than for the exfoliated ones. In the literature, the evolution of the G band frequency with the number of layers, from monolayer graphene to graphite, reveals an up-shift of the G band with the decreasing thickness [24]. The interaction between two graphene layers provokes a slight lattice expansion that leads to the G band frequency down-shift. According to this and in view of the results presented in Table 3, we can diagnose that our pristine few layers samples consist of stacked individual layers with a weak interlayer interaction. Therefore, the initial stress application leads to an increase of the interlayer coupling and probably also to an increase of the substrate-sample coupling; so that the sample is not compressed until the applied stress exceed the 1 GPa threshold. Such effect is more pronounced in the case of CVD samples, since they may contain remnant polymer from the transfer method in between the layers, thereby manifesting a larger interlayer distance in the pristine state than exfoliated samples (Figure S1, Supporting Information). Therefore, we observe that the first stages of compression increases the interlayer coupling, also expelling the polymer out of the sample. The appearance of interlayer shear with stress, as an explanation for the observed G band downshift, can be discarded. We observed that in exfoliated BLG the profile of the 2D band remains unaltered under stress (see Supporting Information), indicating that the AB stacking order is pre-served, thus revealing no shear between graphene layers, as already observed in HOPG samples under similar conditions [22]. In addition, the calculated friction between the planes of the anvil and the disc, even at a highly tilted arrangement, is negligible (Figure S4, Supporting Information). Hence, no or minimal relative movement of the two graphene layers is expected from the FE simulation as well.

Concerning the increase of disorder upon compression, the difference between the exfoliated BLG and the CVD tBLG samples is readily observed by comparison of Figures 1 and 2. In the case of the exfoliated sample, the generation of disorder as a consequence of the high stress treatment is less severe than for the labelled sample. Such a difference is caused by the different quality of the pristine samples; CVD graphene presents an initial degree of disorder, due to the growing and transfer processes, favouring the creation and propagation of defects under stress. For a more detailed analysis, in Figure 4 we present the intensity ratio between the D and the G bands, $I_D/I_G$, as a function of increasing stress. The Raman spectrum of exfoliated BLG before the compression cycle shows no D band (indicating that no defects are generated during the transfer process). The creation of defects with stress starts above ~2.2 GPa and gradually increases up to 0.4 at 3.5 GPa, but $I_D/I_G$ remains constant until the end of the experiment. For the CVD tBLG sample the observed behaviour is clearly different. The uncompressed sample shows some degree of disorder, probably originating during the transfer process (note that in particular for this sample two sequential transfers are performed). During the first stages of compression, up to 1.5 GPa, the intensity of the D band slightly increases from 0.2 to 0.4; but when the applied stress surpasses the 1.5 GPa threshold, the intensity of the D band abruptly increases, becoming almost double the intensity of the G band. From 2.0 GPa the intensity of the D band continues increasing with stress but in a more moderate way, reaching the $I_D/I_G$ intensity ratio of ~2.5 (below the 3.5 threshold for the low to high defects regime [25]). Analogously to HOPG under the same stress conditions [22], the $I_D/I_{D'}$ intensity ratio in tBLG samples remains constant (3.6±0.6) under stress, evidencing the invariance of the prevalent defect type – the edges, i.e. cracks and tears [26] – to the induced stress.

The abrupt increase of $I_D/I_G$ in the tBLG samples is in agreement with the observed G band full width at half maximum (FWHM) evolution with stress (see Figure S5, Supporting Information). In the exfoliated BLG sample the FWHM of the G band follows a linear trend under stress with a slope of ~8.6 cm$^{-1}$/GPa. The FWHM of the G band in the tBLG sample shows the same slope but a discontinuity can be observed between 1.5 and 2.0 GPa, indicating some morphological change in the sample, analogously to what observed in the $I_D/I_G$ evolution. However, it is interesting to note that despite the formation of defects is different for both the exfoliated BLG and the CVD tBLG samples, they present a similar stress behaviour with regards to the stress coefficient of the G band.

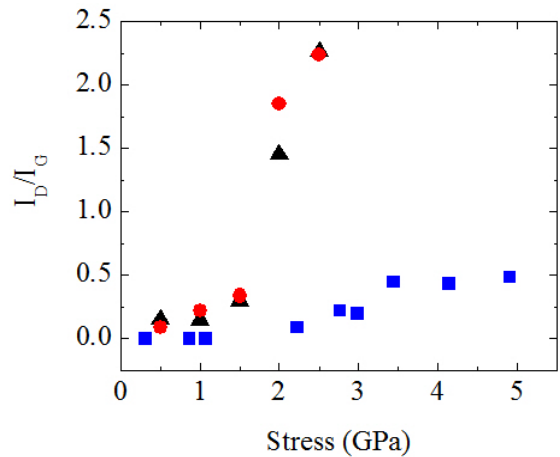

**Figure 4** Intensity ratio between the D and the G bands as a function of stress for exfoliated BLG (blue squares) and for both graphene sheets in labelled tBLG (red circles and black triangles for the $^{13}$C top and $^{12}$C bottom layers, respectively).

**4 Conclusions** In summary, we have presented a high stress study of different graphene samples (exfoliated and CVD, isotopically labelled and of different thickness) in order to address some unknown aspects of the response of graphene to uniaxial out-of-plane stress. While the compressibility of the substrate plays a key role in the high pressure response, reflected in a modified stress coefficient of the G band, the interlayer and layer-substrate coupling effect is only visible at the first stages of compression, up to 1 GPa. Additionally, the doping state of the sample does not seem to have an effect on the mechanical response of graphene - in other words, the high stress shift rates of the Raman bands remain alike for the different specimens regardless the pronounced differences in the initial low stress behaviour. Finally, by comparison of the different studied samples in this work, we can conclude that the generation of defects upon compression does not affect the observed stress coefficient of the G band either.

**Acknowledgements** This work was funded by Czech Science Foundation (project No. 14-15357S) and European Union H2020 Programme (No. 696656 – GrapheneCore1). M.K. and E.d.C. acknowledge the support from ERC-CZ project No. LL1301. M.P.A. is grateful to the European Community for an Internship Erasmus grant.

# SUPPORTING INFORMATION

# Graphene under direct compression: stress effects and interlayer coupling

Elena del Corro, Miriam Peña-Álvarez, Michal Mračko, Radek Kolman, Martin Kalbáč, Ladislav Kavan, Otakar Frank

*Atomic Force Microscopy (AFM)*

The AFM images were obtained using Dimension Icon microscope (Bruker) operating in Peak Force Tapping mode using ScanAsyst-Air probes (stiffness 0.2-0.8 N/m, frequency ~80 kHz). No post-measurement treatment apart from line subtraction (retrace) to remove the tilt has been performed. Figure S1 shows a typical image of a CVD monolayer transferred to sapphire substrate. Minor amount of impurities from the transfer procedure are present, the roughness of the layer (expressed as $R_a$) is ~1.1 nm.

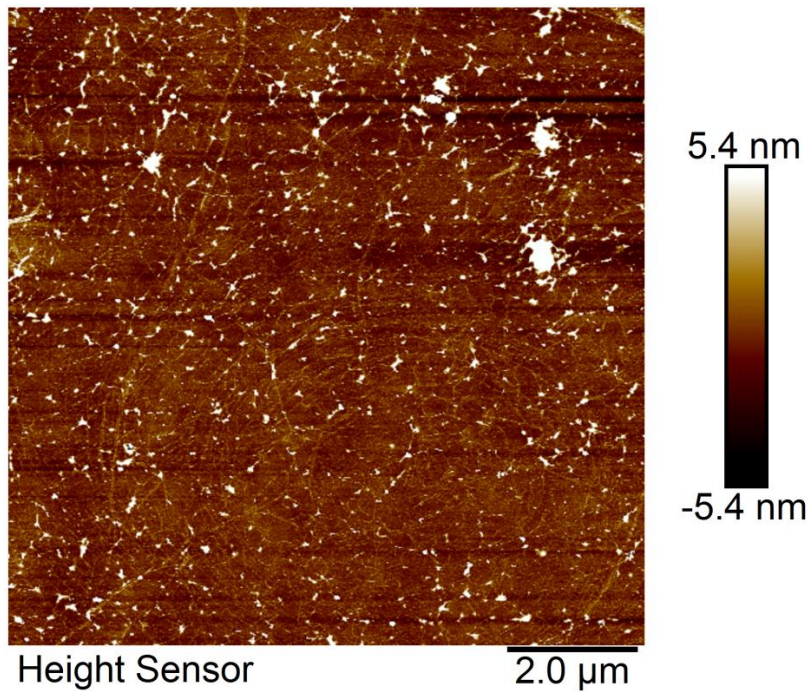

**Figure S1.** AFM image of a graphene CVD monolayer on a sapphire substrate.

*Simulation of the anvil/disc interaction*

The anvil experiment has been modelled using Finite Elements (FE) method. Three configurations have been simulated, all consisting of the sapphire anvil with 350 μm cullet pressing onto a 1 cm (0001) sapphire disc. In one case full parallelism of the anvil and the disc is assumed, in the other cases a tilt angle of 0.1° and 0.15° is considered, respectively (for the calculation details, see below). Figure S2 shows distributions of the contact pressure in the anvil and the disc for parallel and 0.15° tilt.

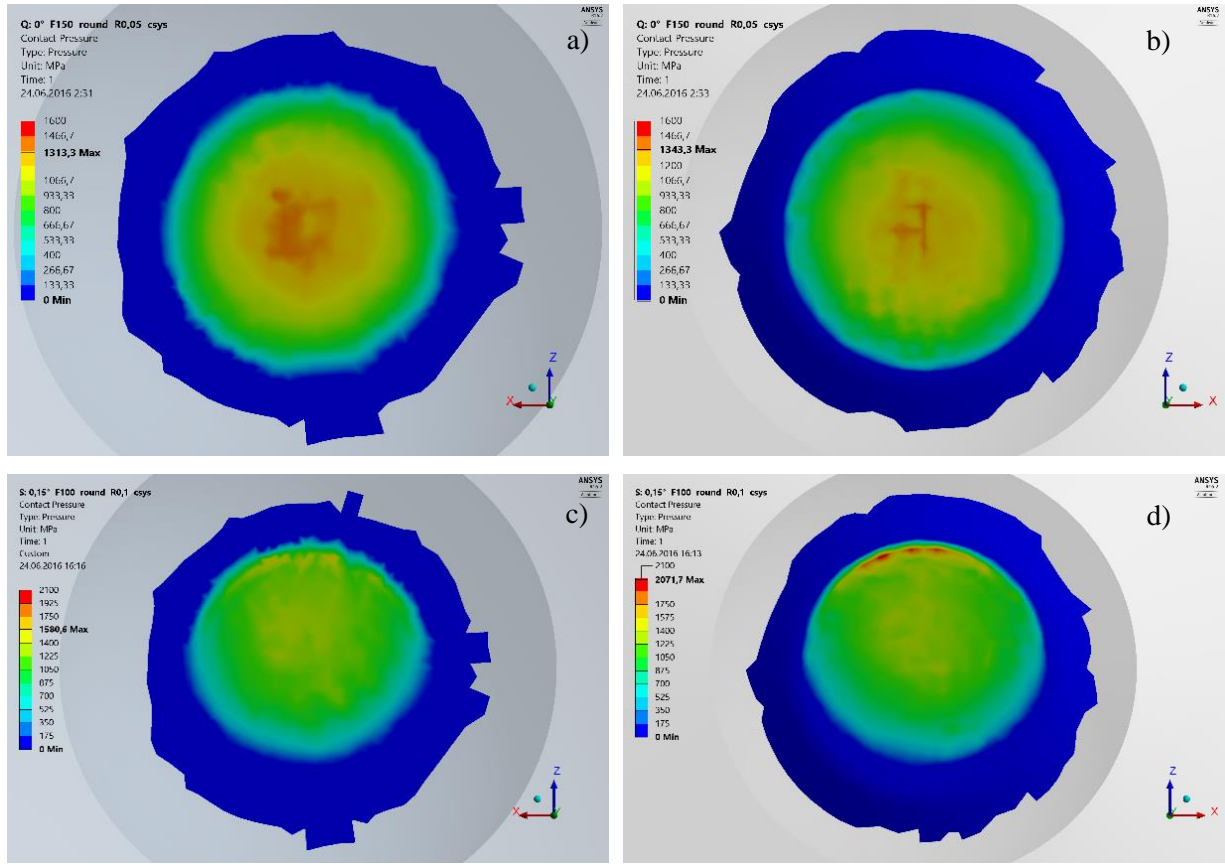

**Figure S2.** Contact pressure distribution in the sapphire disc (a,c) and anvil (b,d) at parallel (top) and 0.15° tilt (bottom) configurations. Note the color scale is the same for both the disc and the anvil in the particular configuration, but different for parallel and tilted arrangement.

Whereas the stress gradient is concentric towards the center in the parallel configuration, in the tilted arrangement the highest stress is reached on the anvil cullet edge and decreases towards the opposite edge. However, for both configurations the stress gradient is negligible in the middle region– where the actual measurements are always conducted. For the flake with lateral dimensions of ~20 μm the maximum edge-to-edge stress difference gives ~120 MPa. Given the stress determination using the sapphire 418 cm$^{-1}$ band with the stress coefficient of 2.1 cm$^{-1}$/GPa [S1] and the point-to-point spectral resolution of ~0.75 cm$^{-1}$ (with 1800 lines/mm grating used for the stress determination), 120 MPa difference is well below the measurement capabilities. Similarly, in graphene, e.g., for the G band with stress coefficient of ~7.5-9.1 cm$^{-1}$/GPa and the point-to-point resolution of ~ 1.8 cm$^{-1}$, the maximum Raman shift difference stemming from the stress gradient at 0.15° tilt is still below the spectral resolution.

The principal stress directions from the FE simulations of the two anvil-disc arrangements are shown in Table S1 (only the values from the experimentally relevant middle element are presented for clarity). In spite of obvious, but nor large, shear components present in the tilted setup (assumed from the direction of the principal components), all the principal stress components are compressive in nature, with the minimum stress direction (i.e., maximum compression) being always normal to the contact plane, and the other two directions are in-plane, perpendicular and close to each other in magnitude. The ratio of the value at the minimum to the value at the maximum (≈ middle) stress direction is approx. 0.55 ± 0.02, in both arrangements, which is in a very good agreement with previous experimental results [S2]. Additionally, there are slight differences between the stresses in the disc and the anvil, however, there is no trend going from parallel to 0.1° and 0.15° tilt, with the variations kept randomly within 1-7%.

**Table S1.** Principal stress directions in the anvil/disc at parallel, 0.1 and 0.15° tilt configuration. Note that for the Euler angles, the initial coordinate system has the XZ plane parallel with the contact plane (Fig. S4d). The ~30° $\theta_{yz}$ shows only the rotation around the vertical axis; without any physical meaning due to the rotational symmetry considerations.

| Angle | Body | Principal Stresses | | | | Euler Angle [°] | | |
|---|---|---|---|---|---|---|---|---|
| | | $\sigma_1$ | $\sigma_2$ | $\sigma_3$ | $\sigma_1/\sigma_3$ | $\theta_{xy}$ | $\theta_{yz}$ | $\theta_{zx}$ |
| 0° | Disc | -694.0 MPa | -755.3 MPa | -1276.9 MPa | 0.54 | -90.8 | 32.5 | -90.0 |
| | Anvil | -739.4 MPa | -804.7 MPa | -1293.6 MPa | 0.57 | -87.6 | 34.9 | 89.8 |
| | Anvil/Disc | 1.07 | 1.07 | 1.01 | | | | |
| 0.1° | Disc | -687.6 MPa | -752.0 MPa | -1291.7 MPa | 0.53 | -92.2 | 30.3 | 93.6 |
| | Anvil | -693.3 MPa | -760.0 MPa | -1253.8 MPa | 0.55 | -91.0 | 35.5 | 87.1 |
| | Anvil/Disc | 1.01 | 1.01 | 0.97 | | | | |
| 0.15° | Disc | -681.5 MPa | -743.6 MPa | -1266.8 MPa | 0.54 | -91.0 | 30.3 | 92.4 |
| | Anvil | -710.7 MPa | -781.2 MPa | -1253.6 MPa | 0.57 | -85.0 | 25.6 | 80.4 |
| | Anvil/Disc | 1.04 | 1.05 | 0.99 | | | | |

Finally, the friction between the disc and the anvil has been calculated for both configurations (Figure S3), clearly showing only a negligible relative movement between the two planes (note the scale in MPa, in contrast to the scale in GPa in Figure S2).

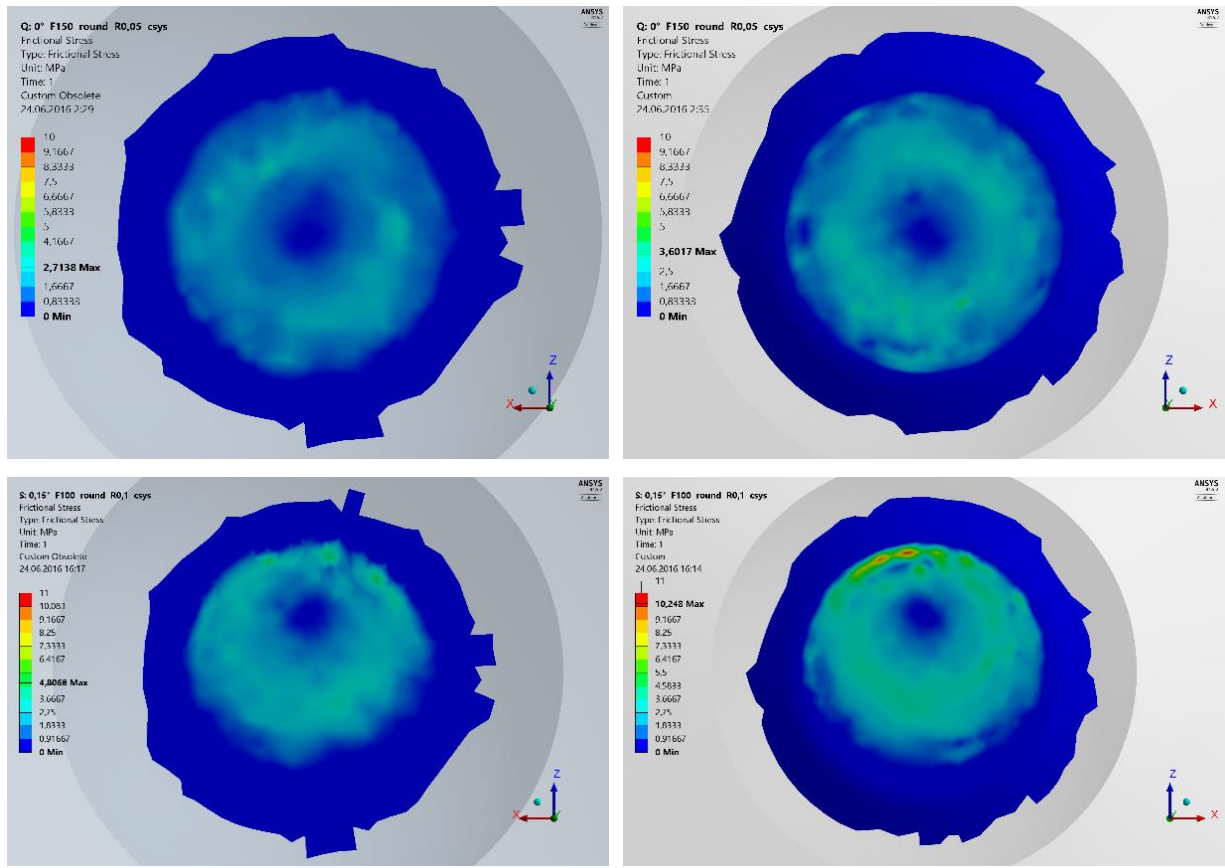

**Figure S3.** Map of friction between the disc (a,c) and the anvil (b, d) calculated for the parallel (top) and the 0.15° tilt (bottom) configurations.

**Simulation method** [S3, S4]. In this Finite Elements analysis a sapphire anvil was pressed against a sapphire disc, considering different anvil/disc orientations (parallel, 0.1° and 0.15° tilted). The geometry of the anvil is shown in Figure S4a; the sapphire disc is a rotational cylinder with diameter of 10 mm and height of 2 mm. Computation could not be performed on ideal geometry because of the singularity on contact surface edge. Therefore, the geometry of the cullet had to be slightly modified, also better representing the real situation. The flat face of the culet was modelled as spherical with a radius of 20 mm and the edge of the cullet was also rounded, with a radius of 0.05 mm. Sapphire is an anisotropic material with trigonal symmetry. In our analysis the c-axis of the crystal is oriented parallel to the rotational axis. The stiffness coefficient in the c-axis direction is represented by $c_{33}$. According to axis-symmetry the rotation of other axes is not important.

$$\begin{pmatrix} c_{11} & c_{12} & c_{13} & c_{14} & 0 & 0 \\ c_{12} & c_{11} & c_{13} & -c_{14} & 0 & 0 \\ c_{13} & c_{13} & C_{33} & 0 & 0 & 0 \\ c_{14} & -c_{14} & 0 & C_{44} & 0 & 0 \\ 0 & 0 & 0 & 0 & C_{44} & c_{14} \\ 0 & 0 & 0 & 0 & c_{14} & 1/2(c_{11}-c_{12}) \end{pmatrix}$$

$c_{11} = 4.902 \; 10^{11}$ Pa  $\qquad c_{44} = 1.454 \; 10^{11}$ Pa  $\qquad c_{13} = 1.130 \; 10^{11}$ Pa

$c_{33} = 4.902 \; 10^{11}$ Pa  $\qquad c_{12} = 1.654 \; 10^{11}$ Pa  $\qquad c_{14} = -0.232 \; 10^{11}$ Pa

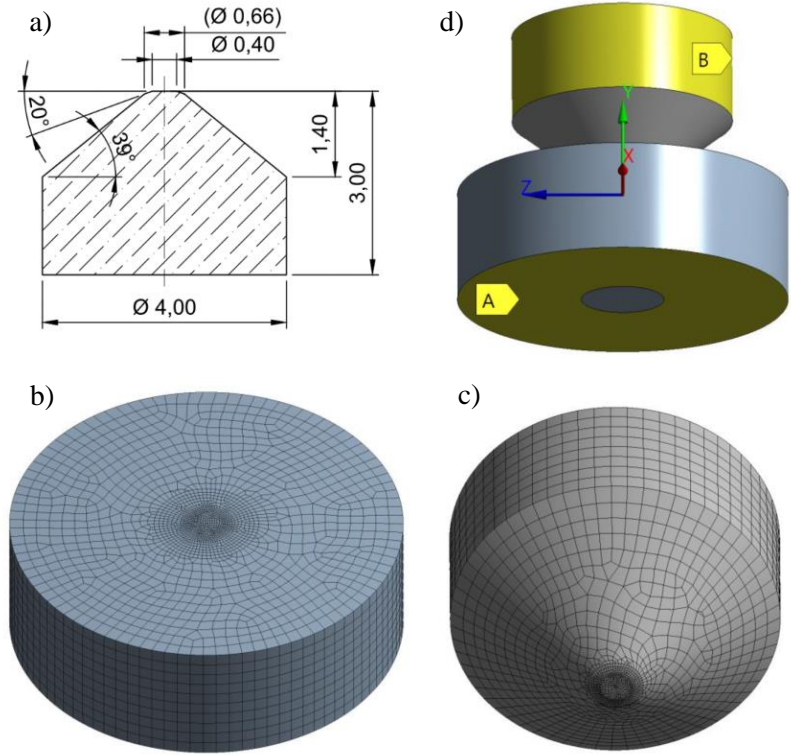

**Figure S4.** a) Geometry and dimensions of the sapphire anvil. b,c) Finite elements composing the disc and the anvil, respectively. d) Opposed anvil/disc configuration.

Unstructured mesh was used for analysis, with quadratic elements. Approximately 60 000 elements (200 000 nodes) consisting of about: 39 000 hexahedrons; 12 000 pyramids; 7 000 tetrahedrons; 2 000 prisms (see Figure S4b-c). Element types used in these analyses can be found as SOLID186 and SOLID187 in Ansys manual.

In the disc, zero displacement was applied in face A, colored in yellow in Figure S4d. Boundary conditions in the anvil were applied with respect to the local coordinate system, where the *z*-axis is oriented along the rotational axis (*c*-axis). In this case, displacement was allowed only in *z*-direction (face B, also in yellow in Figure S4d) and the force was applied on the top face of anvil along the same direction. For contact was used Augmented Lagrange formulation, friction coefficient 0.2 for sapphire to sapphire, with symmetric behavior (no difference between master and slave faces (contact and target in Ansys).

*Full width at half maximum (FWHM) evolution under stress*

For the exfoliated BLG sample the FWHM of the G band follows a linear trend under stress (Figure S5), with a coefficient of 8.6 cm$^{-1}$/GPa. For the CVD tBLG sample we can distinguish two regions, before and after the increase of the defect concentration (Figure 4, main text). As may be expected, a sudden increase of the FWHM of the G band is observed simultaneously to the D band intensity growth. But within both regions, i.e, with low and high defect concentration, the broadening of the spectrum with stress is similar to that obtained for the exfoliated BLG sample; more clearly observed for the $^{13}$C layer (red dots), note that the $^{12}$C G band fitting is more delicate due to the overlapping with the $^{13}$C D' band.

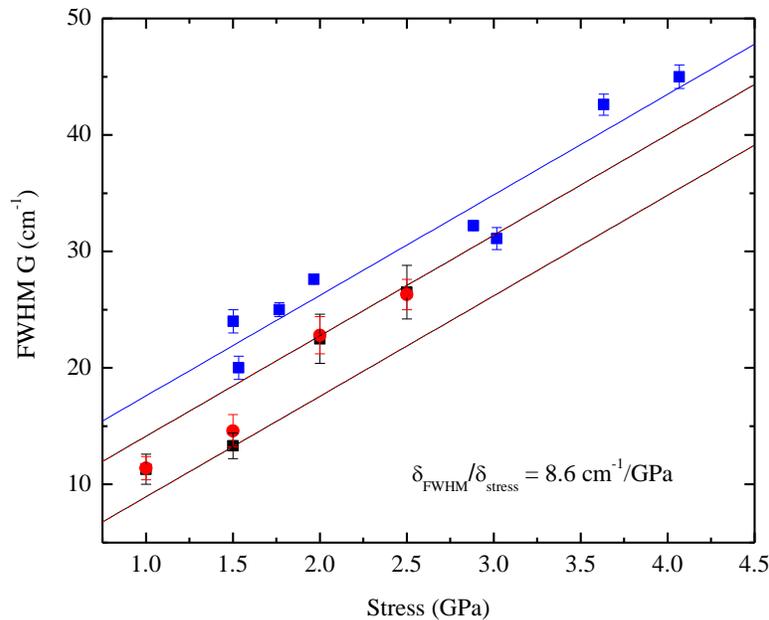

**Figure S5**. FWHM of the G band under stress. Blue dots represent the exfoliated BLG sample and the blue line is the least squares line fit of the experimental data points. Red and black dots represent the CVD tBLG sample, of $^{13}$C and $^{12}$C layers, respectively. Red and black line fits are parallel lines to the blue linear function with 8.6 cm$^{-1}$/GPa slope.

*Absence of interlayer shear under stress*

When few layer graphene samples are directly compressed along the Z axis (perpendicular to the graphene plane) one might expect some shear between the graphene layers; however, such shear was not observed in highly oriented pyrolytic graphite compressed in analogous conditions [S5]. In the case of BLG the appearance of graphene interlayer shear under axial compression is observed neither. The 2D band of pristine BLG shows four contributions, as shown in Figure S6, characteristic of the AB stacking order. We can observe that under 3 GPa compression these four contributions upshift and broaden, but the 2D band profile is kept under stress, indicating that the AB stacking is preserved. Together with the negligible friction (see above), the preservation of the stacking order points to the absence of any interlayer shear.

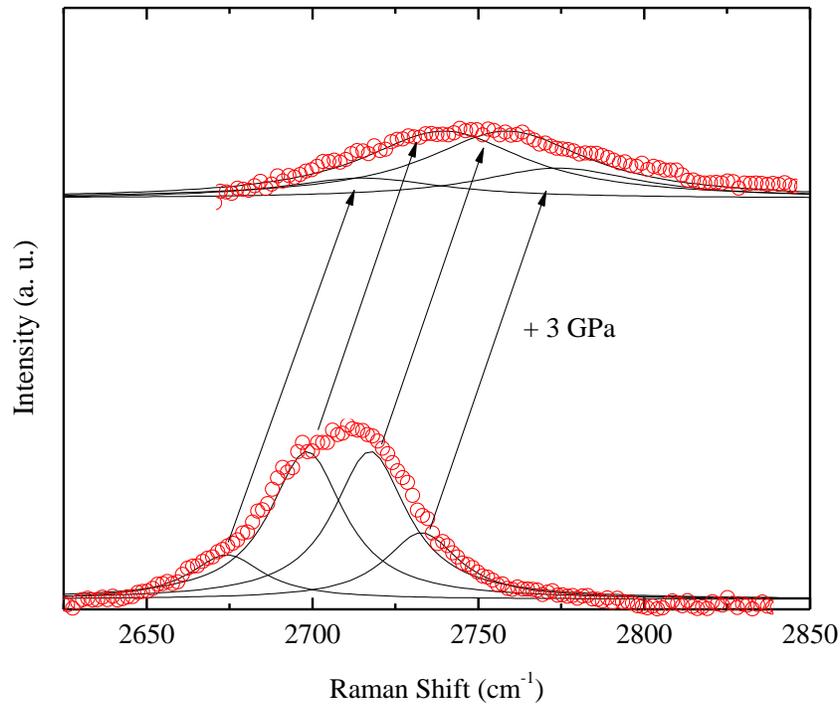

**Figure S6.** Raman spectrum of exfoliated BLG at normal conditions and 3 GPa (red dots for the experimental data points). At normal conditions the black lines represent the Lorentzian fitting of the spectrum; at 3 GPa the black lines represent the simulated Lorentzian contributions obtained from the pristine spectrum line shapes by applying the blue-shift and broadening stress coefficients of this work for each component.